\documentclass{article}

\usepackage{arxiv}

\usepackage[utf8]{inputenc} 
\usepackage[T1]{fontenc}    

\usepackage{algorithm}
\usepackage{algpseudocode}

\usepackage{hyperref}       
\usepackage{url}            
\usepackage{booktabs}       
\usepackage{amsfonts}       
\usepackage{nicefrac}       
\usepackage{microtype}      
\usepackage{graphicx}
\usepackage{natbib}
\usepackage{doi}
\usepackage{multirow}
\usepackage{xcolor}
\usepackage{listings}

\usepackage{amsmath}
\usepackage{amssymb}
\usepackage{mathtools}
\usepackage{amsthm}

\usepackage{wrapfig}

\makeatletter
\AtBeginDocument{}
\makeatother
\newtheorem{definition}{Definition}

\usepackage{amsmath,amsfonts,bm}

\def\eqref#1{equation~\ref{#1}}

\def\1{\bm{1}}

\DeclareMathAlphabet{\mathsfit}{\encodingdefault}{\sfdefault}{m}{sl}
\SetMathAlphabet{\mathsfit}{bold}{\encodingdefault}{\sfdefault}{bx}{n}

\lstdefinestyle{artifact}{
  basicstyle=\ttfamily\scriptsize,
  breaklines=true,
  breakatwhitespace=true,
  columns=fullflexible,
  keepspaces=true,
  frame=single,
  rulecolor=\color{black!30},
  xleftmargin=2pt, xrightmargin=2pt,
  aboveskip=4pt, belowskip=4pt,
  showstringspaces=false,
  upquote=true,
}

\newcommand{\act}[1]{\textsc{#1}}
\newcommand{\scope}[1]{\texttt{#1}}
\newcommand{\Sel}{\mathsf{Select}}
\newcommand{\Exec}{\mathsf{Exec}}
\newcommand{\Gate}{\mathsf{Gate}}
\newcommand{\Refl}{\mathsf{Reflect}}
\newcommand{\Apply}{\mathsf{Apply}}
\newcommand{\Act}{\mathrm{Act}}
\newcommand{\Mem}{\mathcal{M}}

\title{EMMA: Structured Episodic Memory for Multi-Agent Workflow Synthesis}

\author{ {\hspace{0.1mm}Ruijia Niu}\textsuperscript{ 1 2}\thanks{Work done during an internship at Amazon.} \\
	\texttt{rniu@ucsd.edu} \\
	\And{\hspace{0.1mm}Kaiwen Liu}\textsuperscript{ 2}\\
	\texttt{kaiwenl@amazon.com} \\
	\And{\hspace{0.1mm}Rose Yu}\textsuperscript{ 1 2}\thanks{Work performed while Rose Yu was an Amazon Scholar; no longer affiliated with Amazon.}\\
	\texttt{roseyu@ucsd.edu} \\
	\And{\hspace{0.1mm}Kristina Klinkner}\textsuperscript{ 2}\\
	\texttt{klinkner@amazon.com} \\
}

\date{}

\newcommand{\blfootnote}[1]{\begingroup%
\renewcommand\thefootnote{}\footnotetext{#1}%
\addtocounter{footnote}{-1}%
\endgroup}

\renewcommand{\headeright}{}

\renewcommand{\shorttitle}{EMMA: Structured Episodic Memory for Multi-Agent Workflow Synthesis}

\hypersetup{
pdftitle={EMMA: Structured Episodic Memory for Multi-Agent Workflow Synthesis},
pdfauthor={Ruijia Niu, Kaiwen Liu, Rose Yu, Kristina Klinkner},
}

\begin{document}
\maketitle

\setcounter{footnote}{1}
\footnotetext{University of California San Diego}
\setcounter{footnote}{2}
\footnotetext{Amazon}
\blfootnote{Correspondence to: Ruijia Niu <rniu@ucsd.edu>.}
\setcounter{footnote}{0}

\begin{abstract}
Deploying LLM agents in specialized domains often require fine-grained task-specific procedures and knowledge a general model lacks,
yet parameter training is costly and non-interpretable, and hand-engineered agent systems shift
the cost to expert labor. Episodic memory updates at context level, but single agent adaptation struggles to alleviate cross-task interference, especially with complex tasks such as multi-application navigation or data pool analysis. Although multi-agent evolution alleviates this problem, existing methods lack clear definitions on update rules and rely on the backbone model to decide where
and how much the system changes. This \emph{ill-scoped evolution} makes learning
follow the backbone's structural preferences rather than the observed failures: one
backbone accumulates skills while another spawns agents. We present EMMA
(structured Episodic Memory for Multi-Agent workflow synthesis), which treats the
configuration of a multi-agent system (controller routing memory, specialists
inheriting a generalist prompt, and per-step procedural skills) as episodic
memory. EMMA makes the ownership and admissibility of updates explicit: a reflector
assigns each diagnosed failure to the scope owning its correction and proposes one
edit within it, and backbone-independent guards keep every accepted edit
scoped, incremental and bounded. Across KramaBench, AppWorld and BBEH with five
backbones and four baselines (ACE, GEPA, EvoMAS and ADAS), EMMA attains the
best overall performance on every benchmark, up to 32.5\% over the strongest baseline,
maintains consistent improvements across backbones, and uses 17--38\% fewer
training executions than the most efficient baseline.
\end{abstract}

\section{Introduction}
\label{sec:intro}

Deploying LLM agents in specialized domains demands more than general capability.
Real-world applications such as data-to-insight pipelines and multi-application tasks
require tailored conventions and procedures that a general model does not
possess~\citep{kramabench,appworld}. Two mechanisms are conventionally used to supply
such knowledge, and each incurs a distinct cost. Parameter training, whether
supervised fine-tuning or reinforcement learning under objectives such as GRPO and
DAPO~\citep{grpo,dapo}, is computationally expensive and yields an artifact that can
be neither inspected nor selectively corrected. Human engineering of agent systems
avoids that cost but transfers it to expert labor: hand-specified roles and topologies
perform well~\citep{debate,metagpt,smoa} but must be re-authored for every new domain.

Episodic memory offers a more efficient mechanism, in which the experience of solving
a task is persisted as text and reused on later tasks with the model weights held
fixed. Reflexion, A-Mem and MemRL store retrievable reflections, notes and
utility-ranked entries~\citep{reflexion,amem,memrl}, while ACE and GEPA adapt the
context of a ReAct-style agent, as an accumulating playbook and as a reflectively
optimized prompt~\citep{ace,gepa}. Both use a single generic agent
and adapt only its context. Complex applications such as data lake analysis, however,
involve thousands of files from various sources~\citep{kramabench}. A single
agent cannot specialize its behavior per subtask, and its failure becomes observable
only at the end of a long trajectory, so the learning signal cannot be attributed to
the step that caused it.

\begin{figure}[t]
\centering
\includegraphics[width=\linewidth]{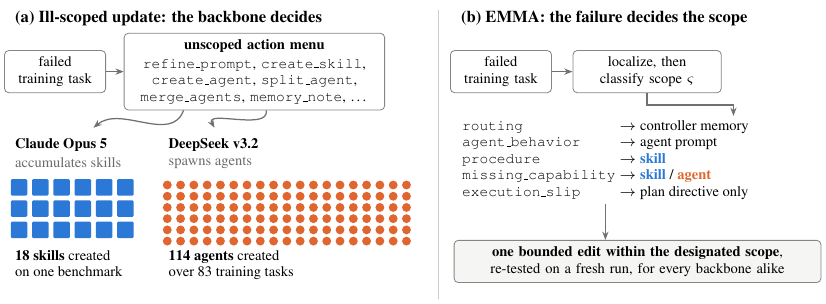}
\vspace{-5mm}
\caption{\textbf{Ill-scoped versus scoped updates.} (a) Without scope constraints, the
update is a backbone habit: Opus~5 accumulates skills, DeepSeek~v3.2 repeatedly creates
agents (one mark per created artifact). (b) EMMA assigns the diagnosed failure to a
scope and makes one bounded edit within it.}
\label{fig:teaser}
\vspace{-3mm}
\end{figure}

Multi-agent evolution and learning systems supply the missing structure. EvoMAS,
EvoAgent, ADAS and AFlow search over agent configurations or programs through
mutation, recombination or meta-agent proposals guided by a task
score~\citep{evomas,evoagentmas,adas,aflow}. In these systems, however, the
optimizing backbone decides \emph{where} and \emph{how extensively} to change the
system: a meta-model proposes a whole new configuration or program, and although
EvoMAS constrains each update step to one component per mutation, there is no limit on
which component is selected, a choice that relies entirely on the backbone model
(Appendix~\ref{app:guard-compare}). This \emph{ill-scoped evolution} makes learning
heavily dependent on the backbone's own behavior. As demonstrated in
Figure~\ref{fig:teaser}, without scope constraints, different backbone models have
almost orthogonal structural preferences, which overshadow the knowledge
actually learned from the task.

We present \textbf{EMMA}, structured \textbf{E}pisodic \textbf{M}emory for
\textbf{M}ulti-\textbf{A}gent workflow synthesis, which makes the \emph{ownership} and
\emph{admissibility} of updates explicit. EMMA treats reusable configurations as
memory derived from task experience. Its controller selects planning knowledge,
agents, and procedural skills, executes one step at a time, and revises unfinished
work from intermediate outputs. Training assigns each diagnosed failure to a scope and
permits a single corresponding edit. Generalization instructions encourage
transferable corrections; programmatic guards enforce scope--action compatibility,
preserve existing prompt content, bound memory growth, and require a change of
strategy after repeated unsuccessful edits. Specialized agents inherit the current
generalist prompt, allowing general improvements to propagate without duplicating
them. Our contributions are:

\begin{itemize}
\item \textbf{Configuration as episodic memory.} We formulate multi-agent workflow
  synthesis as episodic learning over routing knowledge, inheriting agents and
  procedural skills.
\item \textbf{Explicit, guarded updates for consistent update behavior.} Each
  diagnosed failure is assigned to a scope and corrected by a single guarded edit
  within it. This keeps reflection focused on the information a task actually
  provides, instead of superficial structural modifications, repetitive additions, or
  destructive rewrites of existing entries.
\item \textbf{Consistent gains across model families.} EMMA achieves the best mean
  performance on every benchmark, improving over the strongest baseline by up to
  32.5\%, while ranking first in 17 of 20 settings and using the fewest
  training executions.
\end{itemize}

\section{Preliminaries}
\label{sec:prelim}

We consider a distribution $\mathcal{Q}$ over tasks. Each task $q \sim \mathcal{Q}$
is a natural-language request issued against an environment and is paired in
training with a reference $y_q$ (a gold answer or a set of unit tests). A gate
$\Gate(q,\hat{y}) = (v, d)$ returns a binary verdict $v \in \{0,1\}$ for a produced
answer $\hat{y}$ and a short evaluator digest $d$ of why it
failed. $\theta$ denotes the backbone LLM's parameters and
is held \emph{fixed}; the controller, the sub-agents and the reflector are all
instantiated from the same $\theta$.

\subsection{Episodic Learning}
\label{sec:prelim-episodic}

Episodic learning improves an agent by accumulating experience in an external store of
textual entries rather than by updating $\theta$~\citep{reflexion,memrl}. Executing the
agent system on $q$ under memory $\Mem$ is written $\Exec_\theta(\Mem, q) = (\hat{y},
\tau)$, with final answer $\hat{y}$ and trace $\tau$, and learning is a write
$\Mem' = \mathsf{Update}_\theta(\Mem, q, \tau, \Gate(q,\hat y))$.

\begin{definition}[Episodic learning problem]
\label{def:episodic-learning}
Given fixed $\theta$, a training set
$\mathcal{D}_{\mathrm{train}} = \{(q_i, y_{q_i})\}_{i=1}^{N}$ drawn from
$\mathcal{Q}$, and an initial memory $\Mem_0$, form, using only
$\mathcal{D}_{\mathrm{train}}$, a memory
\begin{equation}
\Mem^{\star} \;\in\; \arg\max_{\Mem} \;
\mathbb{E}_{q \sim \mathcal{Q}}\!\left[ v\!\left(q,\, \Exec_\theta(\Mem, q)\right) \right],
\label{eq:objective}
\end{equation}
where $v(q,\cdot)$ is the verdict of the benchmark's official metric. At evaluation
the memory is frozen and $\mathsf{Update}$ is never applied to a held-out task.
\end{definition}

With no gradient, \emph{where} and \emph{how much} $\mathsf{Update}$ changes the store is a design choice.

\begin{figure}[t]
\centering
\includegraphics[width=\linewidth]{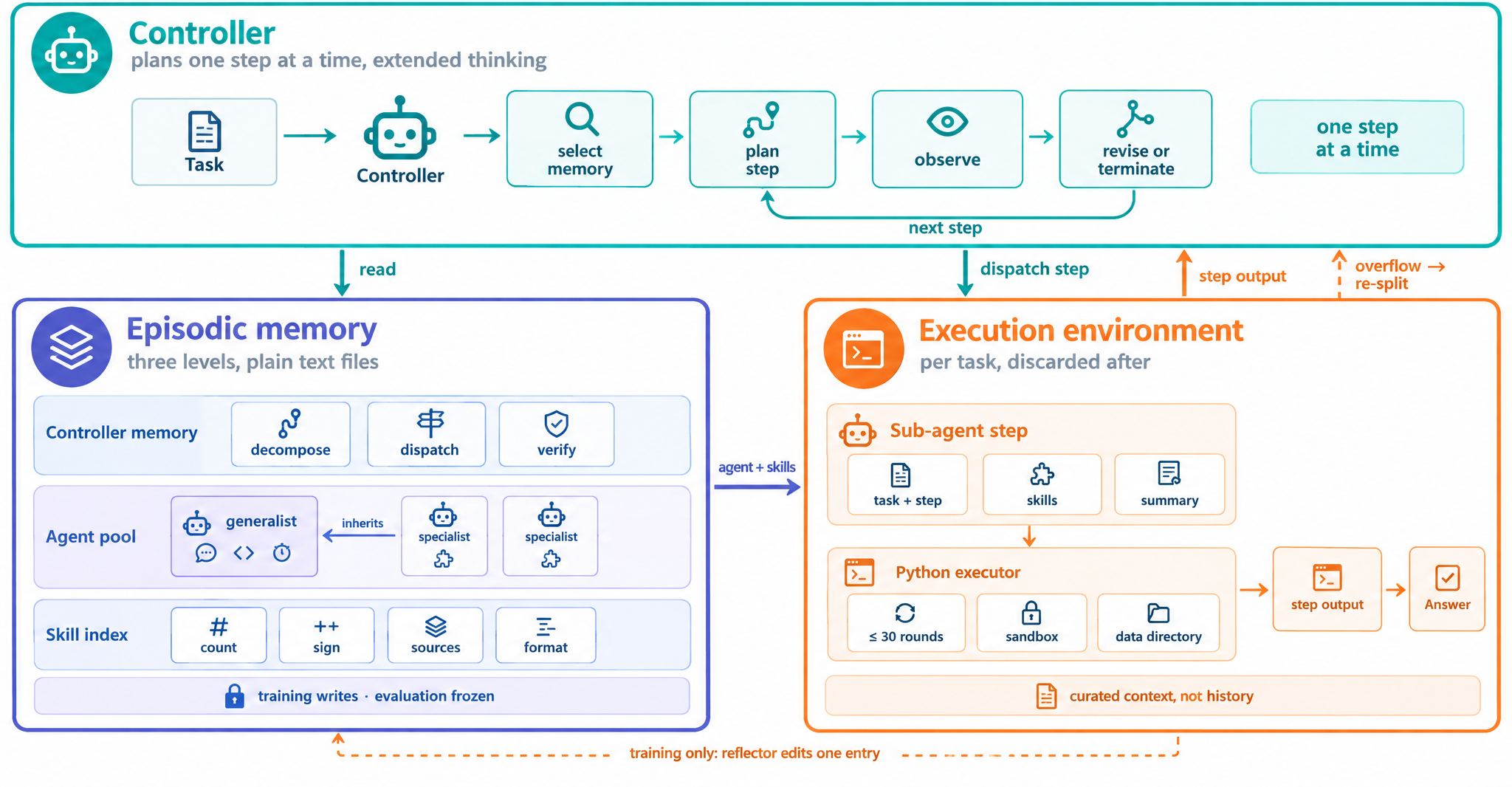}
\vspace{-5mm}
\caption{\textbf{EMMA architecture.} The controller selects controller-memory entries,
plans, and dispatches one step at a time to a sub-agent with its assigned skills;
step outputs drive replanning, and a step that overflows its sub-agent is re-split.
The three memory levels persist across tasks and are written only during training;
the execution environment is created per attempt and discarded.}
\label{fig:architecture}
\vspace{-3mm}
\end{figure}

\subsection{Multi-Agent Workflow with Orchestration}
\label{sec:prelim-workflow}

We adopt the orchestrated setting of Magentic-One, in which a lead agent plans,
tracks progress and re-plans while directing sub-agents that operate the
environment~\citep{magenticone}, and add the per-step assignment of procedural
skills. Let $\mathcal{A}$ be a pool of sub-agents, $\mathcal{S}$ a pool of skills and
$\mathcal{C}$ a controller memory (defined in Section~\ref{sec:method-memory}). A
controller $\pi_c$, itself an LLM with parameters $\theta$, solves $q$ by interleaving
planning with execution. After $t-1$ steps with history
$h_{t-1} = \big((q_1, a_1, S_1, o_1), \dots, (q_{t-1}, a_{t-1}, S_{t-1}, o_{t-1})\big)$,
it emits a subtask, an executing sub-agent, a skill subset and a curated context,
\begin{equation}
\left(q_t,\; a_t,\; S_t,\; \kappa_t\right) \;=\; \pi_c\!\left(q,\, \mathcal{C}_q,\, \mathcal{A},\, \mathcal{S},\, h_{t-1}\right),
\qquad a_t \in \mathcal{A},\;\; S_t \subseteq \mathcal{S},
\label{eq:controller}
\end{equation}
and the sub-agent returns $o_t = \Exec_\theta(a_t, q, q_t, S_t, \kappa_t)$, where
$\mathcal{C}_q \subseteq \mathcal{C}$ is the controller memory loaded for $q$ and
$\kappa_t$ is a controller-authored summary of the prior outputs that step $t$ needs,
rather than the full history. The loop continues until $\pi_c$ terminates and
returns $\hat{y}$.

Because step $t$ is conditioned on $h_{t-1}$, no complete plan is committed in
advance, and because $S_t$ is chosen per step, procedural knowledge is a decision of
the controller rather than a fixed property of a sub-agent. Magentic-One has no skill
layer~\citep{magenticone}, and multi-agent search methods fix $(a_t)_t$ before any $o_t$ is
observed~\citep{evomas,adas,aflow}.

\section{EMMA: Structured Episodic Memory for Multi-Agent Workflow Synthesis}
\label{sec:method}

EMMA instantiates Definition~\ref{def:episodic-learning} with a memory whose entries
\emph{are} the configuration of the multi-agent system of
Section~\ref{sec:prelim-workflow}. Section~\ref{sec:method-arch} describes the
architecture that reads this memory, Section~\ref{sec:method-training} the training
procedure that writes it, and Section~\ref{sec:method-train} summarizes training and
inference. Algorithms~\ref{alg:emma-train}--\ref{alg:emma-reflect} in
Appendix~\ref{app:algorithms} give the exact inputs and outputs of each step.

\subsection{Architecture}
\label{sec:method-arch}

As shown in Figure~\ref{fig:architecture}, EMMA consists of a controller, a persistent
episodic memory, and a task-local execution environment. The memory persists across
tasks and is written only during training; the execution environment is created for
each attempt and discarded afterwards.

\subsubsection{Hierarchical Episodic Memory}
\label{sec:method-memory}

The memory consists of three pools at decreasing levels of generality,
\begin{equation}
\Mem \;=\; \big(\underbrace{\mathcal{C}}_{\text{controller memory}},\;\;
\underbrace{\mathcal{A}}_{\text{agent pool}},\;\;
\underbrace{\mathcal{S}}_{\text{skill pool}}\big).
\label{eq:memory}
\end{equation}
The \textbf{controller memory} $\mathcal{C}$ holds rules about how to decompose a
task, which agent and skills to dispatch, and when to verify. Each entry carries a
topic, a summary, domain tags and a generality label (general, domain-specific or
task-specific). Rather than retrieving by similarity, the controller reads this index
and chooses which entries to load for the current task; loading none is a valid
choice, since irrelevant memory degrades planning.

The \textbf{agent pool} $\mathcal{A}$ holds sub-agents, each defined by a system
prompt $p_j$, an input/output contract and a Python tool. Training starts from a
single generalist $a_0$; every other agent is created during training as a
\emph{specialization} that inherits the generalist's current prompt,
\begin{equation}
\tilde{p}_j \;=\; p_0 \,\Vert\, p_j ,
\label{eq:inherit}
\end{equation}
so a general habit written once in $p_0$ reaches every specialist; $p_j$ holds only
the role-specific method.

The \textbf{skill pool} $\mathcal{S}$ holds short procedures for
\emph{classes} of sub-problems (an algorithm, a rule system, a pitfall checklist,
a format convention), indexed by a one-line ``when to use'' description. The
controller sees only the index and assigns skills per step; the
sub-agent receives their full text.

The three levels answer different questions: $\mathcal{C}$, \emph{how the task is
divided and routed}; the agent prompts, \emph{how an agent works on any subtask}; and
$\mathcal{S}$, \emph{how a class of sub-problem is solved}. This separation gives
every lesson a single home, which the training procedure in \S\ref{sec:method-training} exploits.

\subsubsection{Step-Wise Synthesis with Overflow Decomposition}
\label{sec:method-execution}

Given a task, the controller loads its selected memory and emits an ordered plan in
which each step names an agent, a subtask and the skills to assign. It dispatches one
step at a time (Eq.~\ref{eq:controller}), reads the output, and then continues,
revises the remaining steps, or terminates. Each sub-agent receives the task, its
subtask, its skills and a controller-curated summary of earlier outputs, which keeps
context small while passing on what the step needs.

A step may exhaust its sub-agent, by filling the context window or by running out of
rounds. EMMA reads this as evidence that the subtask was specified too coarsely, not
that the agent failed, and applies \emph{overflow decomposition}: the controller uses
the partial trace to split the unfinished step into smaller steps over existing
agents, splices them into the plan, and resumes at the same position with completed
work preserved. Every overflow is also recorded as evidence for the reflector;
the full procedure is given in Algorithm~\ref{alg:emma-rollout} in
Appendix~\ref{app:algorithms}.

\begin{figure}[t]
\centering
\includegraphics[width=\linewidth]{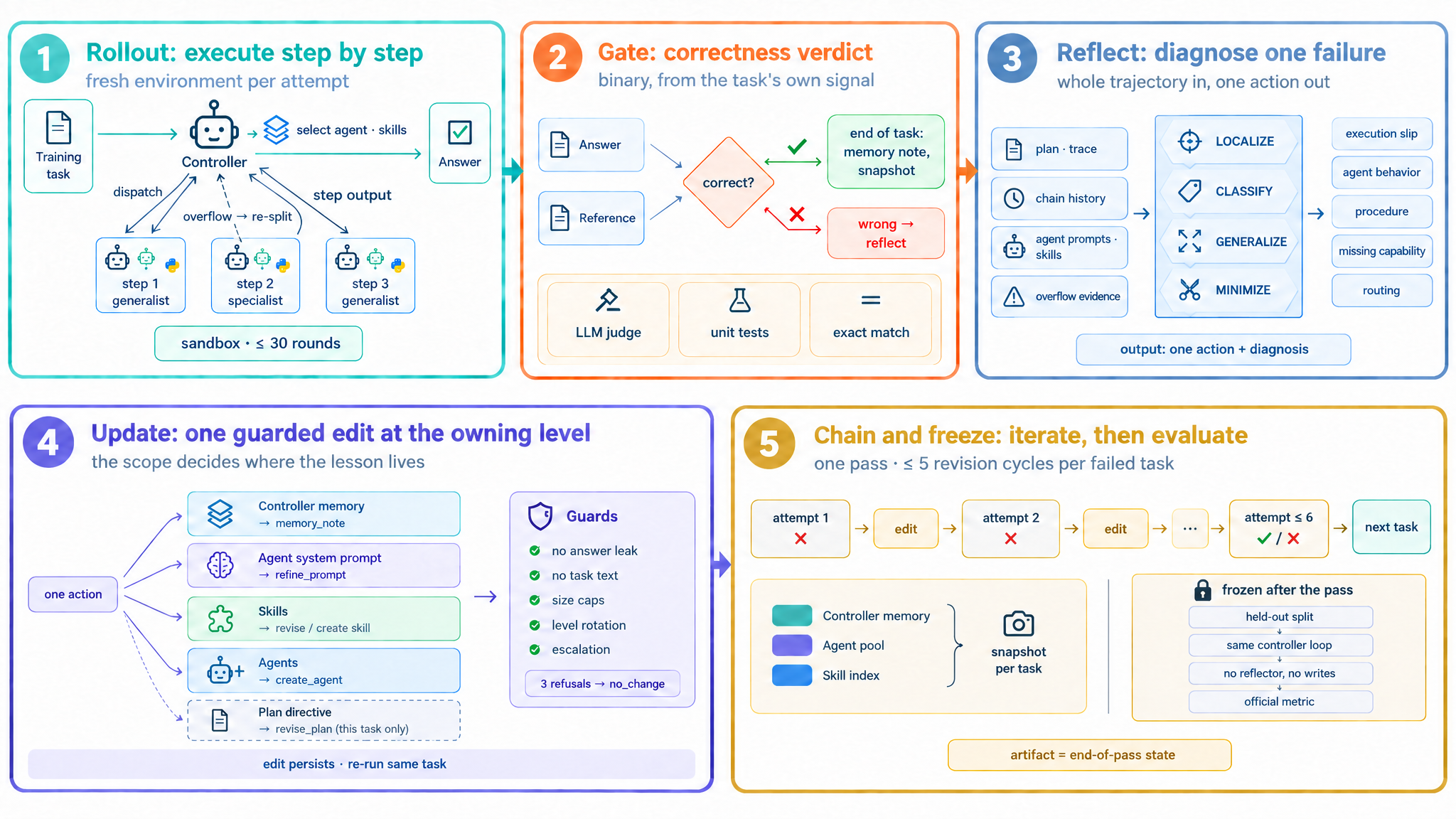}
\vspace{-5mm}
\caption{\textbf{EMMA training procedure.} (1) Rollout in a fresh environment; (2) a
binary gate from the task's own signal; (3) structured reflection over the whole
trajectory (localize, classify, generalize, minimize); (4) one guarded edit within
the designated scope; (5) re-run until success or the revision budget is spent, then
an end-of-task controller update; the memory is frozen after one pass.}
\label{fig:loop}
\vspace{-3mm}
\end{figure}

\subsection{Training Procedure}
\label{sec:method-training}

Figure~\ref{fig:loop} presents the training procedure. Each training task is solved,
gated, diagnosed and revised in a loop, and every revision is one scoped,
guarded edit that is re-tested before the next task.

\begin{table}[t]
\centering
\caption{\textbf{Ownership rules: failure scopes $z$ and available actions}
$\Act(z)$. A lesson at the wrong level never fires again or
contaminates unrelated tasks; the scope--action guard enforces this table.}
\label{tab:scopes}
\scriptsize
\setlength{\tabcolsep}{4pt}
\renewcommand{\arraystretch}{1.1}
\resizebox{\linewidth}{!}{%
\begin{tabular}{llll}
\toprule
\textbf{Scope $z$} & \textbf{Meaning} & \textbf{Level} & \textbf{Available actions $\Act(z)$} \\
\midrule
\scope{execution\_slip}     & knew the method, slipped once              & plan directive (this task) & \act{revise\_plan}, \act{no\_change} \\
\scope{agent\_behavior}     & a general working habit failed             & agent prompt              & \act{refine\_prompt} \\
\scope{procedure}           & method for a class of sub-problems wrong/missing & skill pool          & \act{revise\_skill}, \act{create\_skill} \\
\scope{missing\_capability} & no agent in the pool can do this work      & skills, then agents       & \act{create\_skill}, \act{create\_agent} \\
\scope{routing}             & subtasks correct; decomposition/dispatch wrong & controller memory     & \act{memory\_note}, \act{revise\_plan}; \act{create\_agent} on overflow \\
\bottomrule
\end{tabular}}
\vspace{-3mm}
\end{table}

\subsubsection{Structured Reflection}
\label{sec:method-reflect}

\textbf{Attempt chains.} Training processes each task as an \emph{attempt chain},
shown as steps 1--2 and 5 of Figure~\ref{fig:loop}: the task is executed in a fresh
environment and gated by its own correctness signal (an LLM judge, unit tests or
exact match). A correct answer ends the chain; an incorrect one triggers reflection,
one memory edit and a re-run, up to a fixed number of revisions. No edit is made
after the final failure, since an edit that is never re-run is untested memory.

\textbf{What the reflector sees.} Unlike a scalar fitness, the reflector receives the
whole trajectory of the failed attempt: the plan, the code each sub-agent issued, the
evaluator's explanation of the failure, the history of earlier revisions of this task,
the full prompts of the agents that ran, the assigned skills, and any overflow
events. It may also inspect the full execution log of selected steps.

\textbf{What it decides.} As shown in step 3 of Figure~\ref{fig:loop}, the
reflector follows a fixed four-step procedure. It \emph{localizes} the step where the
outcome went wrong; \emph{classifies} the failure into one of the five scopes listed in
Table~\ref{tab:scopes};
\emph{generalizes} the correction so that it applies to every task of the same kind,
naming no entity or answer of this one, as the reflector's generalization
instructions require; and \emph{minimizes} it to the smallest change
within the lesson's designated scope. It then outputs exactly one action. These
\emph{ownership rules} make the update scoped: each failure scope $z \in \mathcal{Z}$
has a set of available actions $\Act(z)$, which edit only the part of the memory that
owns the correction. A one-off \scope{execution\_slip} is never written to durable memory;
a failed working habit (\scope{agent\_behavior}) is written into a prompt, in $p_0$ if
it is general and in a specialization if it is role-specific; a wrong or missing
method (\scope{procedure}) into a skill; and a wrong decomposition or dispatch
(\scope{routing}) into controller memory. New agents are reserved for work that no
agent can do or that repeatedly overflows one agent.

\subsubsection{Guarded Update}
\label{sec:method-guards}

As shown in step 4 of Figure~\ref{fig:loop}, a proposed action is applied only if
it passes a set of programmatic guards, which depend on the memory, the task and the
revision history but never on the backbone:
\emph{scope--action compatibility}, the action is one of the available actions $\Act(z)$;
\emph{prompt preservation}, a prompt change keeps most of the existing prompt and may
only touch an agent that ran in the failed attempt;
\emph{bounded memory growth}, the agent and skill pools and every entry stay within
fixed sizes, and a new agent must be a genuine specialization rather than a duplicate
or a label;
\emph{no leakage}, no artifact may contain the reference answer or copy a span of the
task text; and
\emph{change of strategy}, no memory level is edited in more than two consecutive
failed revisions, and if the same step overflows twice in a row, the action must
change how the work is split. A refused proposal returns its reason to the reflector,
which tries again; persistent refusals default to no change. A newly created or
revised skill or agent is named in the re-run's plan, so the edit is actually
exercised.

Together, the scoped update and the guards fix the \emph{envelope} of learning: every
accepted edit touches one entry within its designated scope, is incremental, and keeps
the memory bounded, whichever backbone proposed it. The backbone decides \emph{which} available
edit to make, so a stronger model can still learn better memory, but no backbone can
make a larger, misplaced or destructive one, so what is learned follows the training
feedback rather than the backbone's habits. Multi-agent search methods lack such an
envelope, since the meta-model proposes a whole new configuration; as
Figure~\ref{fig:teaser}a shows, when the update is ill-scoped even the \emph{kind} of
edit becomes a backbone habit.

\subsection{Training and Inference}
\label{sec:method-train}

As stated in Algorithm~\ref{alg:emma-train} in Appendix~\ref{app:algorithms}, training makes a single pass over the
training set in a fixed order, running one attempt chain per task; accepted edits persist
regardless of the re-run's outcome. After each chain, one chain-aware controller
update records a few routing lessons, and the memory is snapshotted. After the pass
the memory is frozen, one artifact per benchmark and backbone, with no checkpoint
selection. At inference on held-out tasks, the controller keeps memory selection,
replanning and overflow decomposition, but no reflector is called and nothing is
written.

\section{Related Work}
\label{sec:related}

\textbf{LLM-based multi-agent systems.} Hand-designed systems such as
ChatDev~\citep{qian2024chatdev}, MetaGPT~\citep{metagpt}, AutoGen~\citep{wu2024autogen},
debate and mixture-of-agents topologies~\citep{debate,xu2023towards,smoa} and
Magentic-One~\citep{magenticone} established role specialization and orchestration,
but remain task-specific and carry a coordination
cost~\citep{cemri2025multi,tran2025multi,agenttaxo,agentprune,skillcapacity}.
Automatic design treats the system as code to search, as in ADAS~\citep{adas} and
MAS-GPT~\citep{ye2025mas}; optimizes its structure, as in
GPTSwarm~\citep{zhuge2024gptswarm}, AFlow~\citep{aflow},
AgentSquare~\citep{agentsquare}, AutoAgents~\citep{chen2023autoagents} and generated
topologies~\citep{chen2025argdesigner}; or evolves agents and whole configurations, as
in EvoAgent~\citep{evoagentmas} and EvoMAS~\citep{evomas}. As discussed in Section~\ref{sec:intro}, these methods
let the backbone propose whole candidates, with safeguards for executability or
novelty rather than for the scope of a change (Appendix~\ref{app:guard-compare}),
which makes what they learn sensitive to backbone behavior; EMMA instead admits one
guarded edit within a diagnosed scope.

\textbf{Episodic learning and procedural memory.} Reflexion~\citep{reflexion},
TextGrad~\citep{textgrad}, MIPRO~\citep{mipro} and GEPA~\citep{gepa} improve agents
from natural-language feedback on traces; Dynamic Cheatsheet~\citep{dyncheatsheet},
A-Mem~\citep{amem}, Agent Workflow Memory~\citep{awm}, MemRL~\citep{memrl} and
ACE~\citep{ace} accumulate a store, ACE through incremental curation that
avoids collapse under wholesale rewriting. Skill libraries study how procedures are
authored~\citep{mementoskills,skillrevise}, selected~\citep{skillsinjector} and
routed~\citep{skillrouter,skillcapacity}. These serve a fixed, usually single
ReAct agent~\citep{react}; EMMA stores a multi-agent configuration
and extends incremental updating with a level hierarchy and deterministic
guards.

\section{Experiments}
\label{sec:experiments}

\begin{table}[t]
\centering
\caption{\textbf{KramaBench} (macro accuracy over 76 held-out tasks) and
\textbf{BBEH-mini} (accuracy, 368 held-out examples). Best per column in
\textbf{bold}, second \underline{underlined}; Mean averages over backbones.}
\label{tab:kb-bbeh}
\scriptsize
\setlength{\tabcolsep}{3.2pt}
\renewcommand{\arraystretch}{1.1}
\begin{tabular}{l cccccc c cccccc}
\toprule
& \multicolumn{6}{c}{\textbf{KramaBench}} && \multicolumn{6}{c}{\textbf{BBEH-mini}} \\
\cmidrule(lr){2-7}\cmidrule(lr){9-14}
\textbf{Method} & O-5 & S-5 & GLM-5 & DSv3.2 & K2.5 & \textbf{Mean} && O-5 & S-5 & GLM-5 & DSv3.2 & K2.5 & \textbf{Mean} \\
\midrule
ACE    & 0.694 & 0.545 & 0.602 & 0.478 & 0.579 & 0.580 && \underline{0.780} & 0.514 & 0.353 & 0.492 & 0.448 & 0.517 \\
GEPA   & 0.663 & 0.509 & \underline{0.634} & \underline{0.514} & \textbf{0.602} & \underline{0.584} && 0.764 & \textbf{0.658} & \underline{0.527} & 0.501 & \underline{0.576} & \underline{0.605} \\
EvoMAS & 0.654 & 0.516 & 0.600 & 0.476 & 0.524 & 0.554 && 0.650 & 0.541 & 0.478 & 0.478 & 0.462 & 0.522 \\
ADAS   & \underline{0.709} & \underline{0.584} & 0.530 & 0.481 & 0.504 & 0.562 && 0.690 & 0.595 & 0.500 & \underline{0.516} & 0.519 & 0.564 \\
\midrule
EMMA (ours) & \textbf{0.766} & \textbf{0.641} & \textbf{0.693} & \textbf{0.585} & \underline{0.586} & \textbf{0.654} && \textbf{0.804} & \underline{0.647} & \textbf{0.617} & \textbf{0.563} & \textbf{0.685} & \textbf{0.663} \\
\bottomrule
\end{tabular}
\vspace{-2mm}
\end{table}

\begin{table}[t]
\centering
\caption{\textbf{AppWorld}, TGC\,/\,SGC (\%) on \texttt{test\_normal} (168 tasks) and
\texttt{test\_challenge} (417 tasks). Best in \textbf{bold}, second
\underline{underlined}, ranked per metric. For ADAS on DeepSeek~v3.2 and
Kimi-K2.5, no discovered program beat the hand-written seed, which is reported.}
\label{tab:appworld}
\scriptsize
\setlength{\tabcolsep}{3pt}
\renewcommand{\arraystretch}{1.1}
\begin{tabular}{ll cccccc}
\toprule
\textbf{Split} & \textbf{Method} & O-5 & S-5 & GLM-5 & DSv3.2 & K2.5 & \textbf{Mean} \\
\midrule
\multirow{5}{*}{normal}
 & ACE    & 83.3\,/\,75.0 & \underline{84.5}\,/\,\underline{73.2} & 61.3\,/\,35.7 & \underline{65.5}\,/\,\underline{48.2} & \underline{66.7}\,/\,\underline{35.7} & \underline{72.3}\,/\,\underline{53.6} \\
 & GEPA   & 92.9\,/\,87.5 & 77.4\,/\,55.4 & \underline{63.7}\,/\,\underline{39.3} & 41.7\,/\,7.1 & 53.6\,/\,19.6 & 65.9\,/\,41.8 \\
 & EvoMAS & \underline{94.6}\,/\,\underline{89.3} & 69.0\,/\,44.6 & 16.7\,/\,1.8 & 31.5\,/\,10.7 & 44.6\,/\,30.4 & 51.3\,/\,35.4 \\
 & ADAS   & 90.5\,/\,83.9 & 67.9\,/\,53.6 & 22.0\,/\,12.5 & 0.0\,/\,0.0 & 3.6\,/\,0.0 & 36.8\,/\,30.0 \\
 & EMMA (ours) & \textbf{96.4}\,/\,\textbf{94.6} & \textbf{89.9}\,/\,\textbf{82.1} & \textbf{78.6}\,/\,\textbf{57.1} & \textbf{77.4}\,/\,\textbf{57.1} & \textbf{78.6}\,/\,\textbf{64.3} & \textbf{84.2}\,/\,\textbf{71.0} \\
\midrule
\multirow{5}{*}{challenge}
 & ACE    & 88.0\,/\,\underline{81.3} & \underline{80.6}\,/\,\underline{65.5} & \underline{56.1}\,/\,\underline{36.0} & \underline{46.0}\,/\,\underline{24.5} & \textbf{66.2}\,/\,\textbf{46.0} & \underline{67.4}\,/\,\underline{50.7} \\
 & GEPA   & 85.1\,/\,71.9 & 70.5\,/\,49.6 & 44.1\,/\,23.0 & 23.5\,/\,5.0 & 41.2\,/\,12.2 & 52.9\,/\,32.3 \\
 & EvoMAS & \underline{89.0}\,/\,78.4 & 52.3\,/\,36.0 & 7.2\,/\,0.0 & 11.3\,/\,1.4 & 24.5\,/\,10.1 & 36.9\,/\,25.2 \\
 & ADAS   & 84.4\,/\,71.9 & 36.2\,/\,17.3 & 7.7\,/\,2.9 & 0.5\,/\,0.0 & 1.2\,/\,0.0 & 26.0\,/\,18.4 \\
 & EMMA (ours) & \textbf{89.2}\,/\,\textbf{83.5} & \textbf{84.7}\,/\,\textbf{73.4} & \textbf{71.5}\,/\,\textbf{51.1} & \textbf{62.6}\,/\,\textbf{40.3} & \underline{61.2}\,/\,\underline{31.6} & \textbf{73.8}\,/\,\textbf{56.0} \\
\bottomrule
\end{tabular}
\vspace{-3mm}
\end{table}

\subsection{Experimental Setup}
\label{sec:exp-setup}

\textbf{Benchmarks.} \textbf{KramaBench}~\citep{kramabench} (28 train / 76 held-out):
data extraction, integration and analysis across six scientific domains; macro
accuracy over domains. \textbf{AppWorld}~\citep{appworld} (90 train / 168
\texttt{test\_normal} / 417 \texttt{test\_challenge}): tasks across nine applications
through APIs, where challenge tasks need an application unseen in training; task-goal
(TGC) and scenario-goal completion (SGC, all variants of a scenario must pass) from
state-based unit tests. \textbf{BBEH-mini}~\citep{bbeh} (92 train / 368 held-out): 23
equally sized reasoning categories; accuracy. Splits are disjoint and shared by all
methods (Appendix~\ref{app:details}).

\textbf{Backbones and baselines.} Claude Opus~5~\citep{claudeopus5} and
Sonnet~5~\citep{claudesonnet5}, and the open-weight GLM-5~\citep{glm5},
DeepSeek~v3.2~\citep{deepseekv32} and Kimi-K2.5~\citep{kimik25}; within a run the controller, sub-agents and
reflector share one backbone. We compare with four methods that learn persistent
artifacts from the same training split with ground-truth feedback and freeze them
(Appendix~\ref{app:baselines}): \textbf{ACE}~\citep{ace}, a curated playbook for a
fixed ReAct agent; \textbf{GEPA}~\citep{gepa}, reflective prompt evolution;
\textbf{EvoMAS}~\citep{evomas}, evolution of whole multi-agent configurations rewarded
by the dataset evaluator minus a cost term; and \textbf{ADAS}~\citep{adas}, meta-agent
search over agent programs with a budget matched to GEPA. EMMA's budgets, caps and
guard settings are listed in Appendix~\ref{app:config}.

\begin{figure}[t]
\centering
\includegraphics[width=\linewidth]{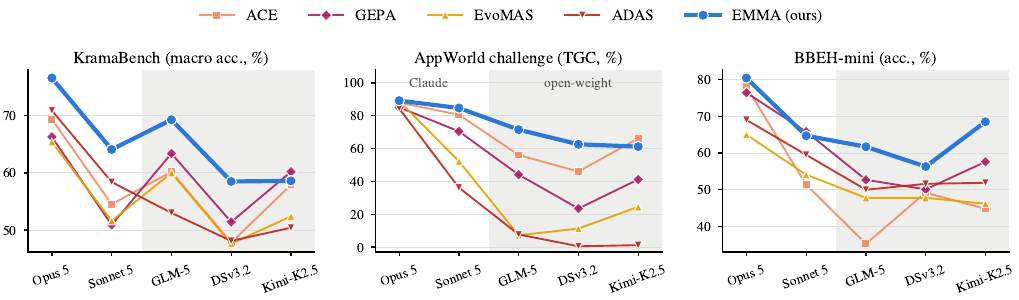}
\vspace{-5mm}
\caption{\textbf{Cross-family consistency.} Held-out performance per backbone (from
Tables~\ref{tab:kb-bbeh} and~\ref{tab:appworld}); shaded columns are open-weight
backbones.}
\label{fig:crossfamily}
\vspace{-3mm}
\end{figure}

\subsection{Experimental Results}
\label{sec:exp-main}

As shown in Tables~\ref{tab:kb-bbeh} and~\ref{tab:appworld}, EMMA attains the
\textbf{highest mean on all three benchmarks}: $+7.0$ accuracy points over the
strongest baseline on KramaBench (0.654 vs.\ 0.584, GEPA), $+5.8$ on BBEH-mini (0.663
vs.\ 0.605, GEPA), and $+11.9$ TGC\,/\,$+17.4$ SGC on AppWorld normal and $+6.4$
TGC\,/\,$+5.3$ SGC on challenge (over ACE). Counting the AppWorld splits separately,
EMMA ranks \textbf{first in 17 of 20 benchmark--backbone settings} and second in the
other three (KramaBench and AppWorld challenge on Kimi-K2.5, BBEH on Sonnet~5); TGC and
SGC give the same rankings. The runner-up changes with the benchmark (GEPA on
KramaBench and BBEH, ACE on AppWorld), so the advantage does not rest on one weak
comparison. The largest gains are on AppWorld challenge with open-weight backbones
($+16.6$ TGC on DeepSeek~v3.2, $+15.4$ on GLM-5).

\subsubsection{Consistency Across Model Families}
\label{sec:exp-consistency}

Figure~\ref{fig:crossfamily} shows how each method's held-out performance changes
across backbones, and reveals three regularities.
\textbf{(i) Multi-agent search methods are the most backbone-sensitive.} On AppWorld
challenge, EvoMAS and ADAS approach EMMA with Opus~5 (89.0 and 84.4 vs.\ 89.2 TGC) but
fall to 7.2 (EvoMAS, GLM-5) and 0.5 (ADAS, DeepSeek~v3.2); on two open-weight backbones
ADAS finds no program better than its hand-written seed. When the backbone authors the
whole configuration, the result is only as good as that authorship.
\textbf{(ii) EMMA's weakest backbone is stronger than every baseline's weakest
backbone} on every benchmark (58.5 on KramaBench, 61.2 TGC on AppWorld challenge and
56.3 on BBEH), and on KramaBench and AppWorld normal it even exceeds every baseline's
five-backbone mean.
\textbf{(iii) EMMA's margin grows on open-weight backbones} on AppWorld and BBEH: over
the strongest baseline per backbone, it leads by $+0.2$ TGC with Opus~5 but $+15.4$ and
$+16.6$ with GLM-5 and DeepSeek~v3.2 on AppWorld challenge, and by $+2.4$ with Opus~5
but $+9.0$ and $+10.9$ with GLM-5 and Kimi-K2.5 on BBEH. On KramaBench all methods
degrade similarly across families, and EMMA's lead is a steady shift of about six
points on four of the five backbones.

\begin{table}[t]
\centering
\begin{minipage}[t]{0.40\linewidth}
\centering
\caption{\textbf{Training task executions}, mean over five backbones (lower is
better). Best in \textbf{bold}, second \underline{underlined}.}
\label{tab:efficiency}
\scriptsize
\renewcommand{\arraystretch}{1.62}
\begin{tabular*}{\linewidth}{@{\extracolsep{\fill}}lccc@{}}
\toprule
\textbf{Method} & \textbf{KB} & \textbf{BBEH} & \textbf{AppW.} \\
\midrule
ACE    & \underline{96.8} & \underline{192.6} & 283.4 \\
GEPA   & 156.4 & 222.8 & 229.8 \\
EvoMAS & 280.0 & 385.1 & 1{,}098.0 \\
ADAS   & 141.4 & 200.0 & \underline{206.0} \\
\midrule
EMMA (ours) & \textbf{61.8} & \textbf{159.6} & \textbf{127.8} \\
\bottomrule
\end{tabular*}
\end{minipage}\hfill
\begin{minipage}[t]{0.56\linewidth}
\centering
\caption{\textbf{Ablations on KramaBench with Opus~4.5.} Top: learned memory
levels. Bottom: training-split episodic learning vs.\ test-split adaptation.}
\label{tab:ablation}
\scriptsize
\renewcommand{\arraystretch}{1.1}
\begin{tabular*}{\linewidth}{@{\extracolsep{\fill}}llc@{}}
\toprule
\multicolumn{2}{@{}l}{\textbf{Configuration}} & \textbf{Acc.} \\
\midrule
\multicolumn{2}{@{}l}{EMMA, agent and skill pools untrained} & 0.506 \\
\multicolumn{2}{@{}l}{EMMA, fixed agent pool (skills + ctrl.\ memory)} & \underline{0.591} \\
\multicolumn{2}{@{}l}{EMMA (ours), full} & \textbf{0.647} \\
\midrule
Within-task reflection (test) & DACode & 0.443 \\
 & Reflexion & 0.476 \\
 & DeepResearch & \underline{0.535} \\
Cross-task self-evolution (test) & EvoMAS & 0.532 \\
Cross-task episodic (train) & EMMA (ours) & \textbf{0.647} \\
\bottomrule
\end{tabular*}
\end{minipage}
\vspace{-3mm}
\end{table}

\subsubsection{Update Efficiency}
\label{sec:exp-efficiency}

As reported in Table~\ref{tab:efficiency}, EMMA achieves the best mean performance with
the \textbf{fewest training executions on every benchmark}, 17--38\% fewer than the most
efficient baseline. EMMA re-runs only failed tasks, and each re-run tests one localized,
guarded edit. This result demonstrates that these updates are on-point: each training
execution yields a targeted correction rather than a speculative restructuring, which
maximizes learning efficiency over the baselines.

\subsection{Ablation Studies}
\label{sec:exp-ablation}

All ablations are conducted on KramaBench with Claude Opus~4.5~\citep{claudeopus45},
under which full EMMA attains 0.647; Table~\ref{tab:ablation} reports all compared
configurations.

\textbf{Memory levels.} This ablation asks whether treating the multi-agent
configuration as memory contributes beyond what a fixed system can learn. Controller
memory is learned in every arm, while the skill and agent pools are either kept at
their initial state or learned. Learning procedural skills accounts for the larger of the
two gains, which indicates that much of the transferable knowledge concerns how
recurring classes of sub-problems are solved. Learning the agent pool as well brings a
further gain (0.591 to 0.647), showing that specialized agents capture knowledge that
skills alone do not, and supporting our premise that the structure of the system,
not only its context, should be learned. Because controller memory is present in all
arms, its individual contribution is not isolated.

\textbf{Episodic learning vs.\ test-time adaptation.} This ablation asks whether the
gains stem from memory learned on training tasks rather than from adaptation on the
evaluated tasks themselves. We compare EMMA with within-task reflection (DACode,
Reflexion and smolagents DeepResearch;~\citealp{dacode,reflexion,smolagents}), which
revises a solution within one trajectory, and with EvoMAS applied as cross-task
self-evolution on the test split without ground truth. All test-time variants remain
well below EMMA, indicating that corrections derived from verified training failures
transfer to held-out tasks more effectively than additional adaptation at test time.
As these arms also differ in feedback access and adaptation split, the comparison
supports the transfer of learned memory rather than isolating persistence alone.

\textbf{Case study: what training produces.} In the Sonnet~5 run on BBEH, a single
failed zebra puzzle led to three artifacts, each written at the level that owns it: a
constraint-satisfaction skill, a specialist agent created from overflow evidence, and
a routing rule in controller memory. Subsequent zebra tasks were solved on the first
attempt, and a held-out puzzle was solved and verified unique in a single dispatched
step (Appendix~\ref{app:artifacts}). This illustrates how scoped updates distribute
one lesson across the memory hierarchy and how the resulting artifacts are reused.
\label{sec:exp-qualitative}

\section{Conclusion and Limitations}
\label{sec:conclusion}

In this paper, we present EMMA, structured Episodic Memory for Multi-Agent workflow
synthesis, which treats the configuration of a multi-agent system (controller routing
memory, inheriting agents and procedural skills) as episodic memory and updates it
only within the scope of each diagnosed failure, through a single edit that passes
backbone-independent guards. The experimental results show that EMMA achieves the
highest mean performance on every benchmark and split across five backbones from two
model families, ranks first in 17 of 20 benchmark--backbone settings, and uses the
fewest training executions, while the multi-agent search baselines degrade sharply on
open-weight backbones. These results indicate that scoped, guarded updates let what
is learned follow the training feedback rather than the backbone's structural
preferences.

Despite these advantages, EMMA narrows but does not eliminate the performance gap
between backbone families, and pairing a strong reflector with weaker executor agents
is a natural way to exploit this. In addition, accepted edits persist even when an
attempt chain ends unsolved; we observed no systematic degradation, but ineffective or
harmful entries may remain, and cross-verification on mini-batches or multiple epochs
could identify them. We consider both directions as future work.

\newpage

\bibliography{emma_refs}
\bibliographystyle{unsrtnat}

\newpage
\appendix
\section{Algorithms}
\label{app:algorithms}

Algorithm~\ref{alg:emma-train} states the training loop of Figure~\ref{fig:loop}. Algorithm~\ref{alg:emma-rollout}
states $\textsc{Rollout}$, which executes one attempt under the current memory, and
Algorithm~\ref{alg:emma-reflect} states $\textsc{ReflectGuarded}$, which produces one
available action for a failed attempt. Notation: $a_0$ is the generalist and
$\tilde{p}_j$ an agent's effective prompt (Eq.~\ref{eq:inherit}); a plan
$\Pi=(\pi_1,\dots,\pi_m)$ has steps $\pi_t=(a_t,q_t,S_t)$ (agent, subtask, skills);
$\mathrm{idx}(\cdot)$ denotes an index of topics or ``when to use'' descriptions;
$\omega_t$ records whether step $t$ overflowed its context or round budget; $\eta$ is a
task-local plan directive; $(v,d)$ is the gate's verdict and evaluator digest; and
$(z,\alpha,\delta)$ are the reflector's scope, action and payload.

\begin{algorithm}[h]
\caption{\textbf{EMMA training.} Only the memory $\Mem = (\mathcal{C},\mathcal{A},\mathcal{S})$
changes; $\theta$ is fixed. One pass, one attempt chain per task, at most one guarded edit per
failed attempt.}
\label{alg:emma-train}
\begin{algorithmic}[1]
\small
\Require training tasks $\mathcal{D}_{\mathrm{train}}=\{(q_i,y_i)\}_{i=1}^{N}$; seed memory
         $\Mem_0 = (\varnothing, \{a_0\}, \varnothing)$; revisions $K$
\Ensure  frozen memory $\Mem$
\State $\Mem \gets \Mem_0$
\For{$i = 1$ \textbf{to} $N$} \Comment{fixed order, single pass}
  \State $H \gets \langle\,\rangle$;\; $\eta \gets \varnothing$
  \For{$k = 0$ \textbf{to} $K$}
    \State $(\hat{y},\tau,\Omega) \gets \textsc{Rollout}(q_i,\Mem,\eta)$ \Comment{fresh environment; Alg.~\ref{alg:emma-rollout}}
    \State $(v,d) \gets \Gate(q_i,\hat{y})$
    \If{$v = 1$ \textbf{or} $k = K$} \textbf{break} \Comment{no untested revision after the last failure} \EndIf
    \State $x \gets (q_i, d, \Pi, \tau_{\mathrm{code}}, H, \{\tilde p_j\}_{a_j\in\tau}, \mathrm{idx}(\mathcal{S}), S_\tau, \Omega)$ \Comment{reflection input; \S\ref{sec:method-reflect}}
    \State $(\alpha,\delta,z) \gets \textsc{ReflectGuarded}(x)$ \Comment{Alg.~\ref{alg:emma-reflect}}
    \State $\Mem \gets \Apply(\alpha,\delta,\Mem)$;\; $\eta \gets \textsc{Directive}(\alpha,\delta)$;\; $H \gets H \oplus (z,\alpha,v)$
  \EndFor
  \State $\mathcal{C} \gets \textsc{UpdateController}_\theta(\mathcal{C}, q_i, H, v)$ \Comment{$\le 2$ entries, pass or fail}
  \State $\textsc{Snapshot}(\Mem)$
\EndFor
\State \Return $\Mem$ \Comment{frozen for evaluation}
\end{algorithmic}
\end{algorithm}

\begin{algorithm}[h]
\caption{$\textsc{Rollout}$: controller-mediated execution of one attempt. The plan is mutable;
a step that overflows its sub-agent is re-split and re-entered at the same index.}
\label{alg:emma-rollout}
\begin{algorithmic}[1]
\small
\Require task $q$; memory $\Mem=(\mathcal{C},\mathcal{A},\mathcal{S})$; directive $\eta$;
         caps $K_\Pi$ (plan steps), $R$ (rounds per step), $R_{\mathrm{re}}$ (re-plans)
\Ensure  answer $\hat{y}$; trace $\tau$; overflow evidence $\Omega$
\State $\mathcal{C}_q \gets \Sel_\theta(q,\mathrm{idx}(\mathcal{C}))$ \Comment{controller-selected; $\varnothing$ allowed}
\State $\Pi \gets \textsc{Plan}_\theta(q,\mathcal{C}_q,\mathrm{idx}(\mathcal{A}),\mathrm{idx}(\mathcal{S}),\eta)$ \Comment{$\pi_t=(a_t,q_t,S_t)$}
\State $\tau \gets \langle\,\rangle$;\; $\Omega \gets \varnothing$;\; $t \gets 1$
\While{$t \le |\Pi|$ \textbf{and} $t \le K_\Pi$}
  \If{$a_t \notin \mathcal{A}$} \State $\Pi \gets \textsc{Replan}_\theta(\Pi, t)$; \textbf{continue} \Comment{strict pool at training; $\le R_{\mathrm{re}}$ times} \EndIf
  \State $\kappa_t \gets \textsc{Curate}_\theta(\tau, \pi_t)$
  \State $o_t \gets \Exec_\theta(\tilde{p}_{a_t}, q, q_t, S_t, \kappa_t)$ \Comment{$\le R$ rounds of Python; Eq.~\ref{eq:inherit}}
  \If{$\omega_t \neq \varnothing$} \Comment{context or round overflow}
    \State $\Omega \gets \Omega \cup \{(t, a_t, \omega_t, q_t, \mathrm{partial}(o_t))\}$
    \State $\Pi' \gets \textsc{Decompose}_\theta(\pi_t,\mathrm{partial}(o_t),\mathcal{A},\mathcal{S})$
    \If{$\Pi' \ne \varnothing$} \State $\Pi \gets \Pi_{1:t-1} \oplus \Pi' \oplus \Pi_{t+1:|\Pi|}$; \textbf{continue} \Comment{splice, re-enter at $t$} \EndIf
  \EndIf
  \State $\tau \gets \tau \oplus (\pi_t,o_t)$
  \State $\textit{dec} \gets \textsc{Decide}_\theta(q, \tau, \Pi_{t+1:})$ \Comment{continue / revise remaining steps / terminate}
  \If{$\textit{dec} = \textsc{terminate}$} \textbf{break} \ElsIf{$\textit{dec} = \textsc{revise}$} \State $\Pi \gets \Pi_{1:t} \oplus \textsc{Revise}_\theta(q,\tau)$ \EndIf
  \State $t \gets t+1$
\EndWhile
\State \Return $\textsc{FinalAnswer}_\theta(\tau),\ \tau,\ \Omega$
\end{algorithmic}
\end{algorithm}

\begin{algorithm}[h]
\caption{$\textsc{ReflectGuarded}$: one available action for a failed attempt.}
\label{alg:emma-reflect}
\begin{algorithmic}[1]
\small
\Require reflection input $x$: task, digest, plan, per-round code $\tau_{\mathrm{code}}$, chain history $H$, prompts of participating agents, skill-pool index and injected skills $S_\tau$, overflow evidence $\Omega$
\Ensure  action $\alpha$, payload $\delta$, scope $z$
\State $\textit{fb} \gets \varnothing$
\For{$r = 1$ \textbf{to} $3$}
  \State $(t^\star,z,\alpha,\delta) \gets \Refl_\theta(x, \textit{fb})$ \Comment{localize, classify, generalize, minimize; may inspect logs}
  \State $\textit{why} \gets \textsc{FirstViolatedGuard}(z,\alpha,\delta,\Mem,H,\Omega)$ \Comment{order in Appendix~\ref{app:config}}
  \If{$\textit{why} = \varnothing$} \State \Return $(\alpha,\delta,z)$ \EndIf
  \State $\textit{fb} \gets \textit{fb} \oplus \textit{why}$ \Comment{refusal reason fed back}
\EndFor
\State \Return $(\act{no\_change}, \varnothing, z)$
\end{algorithmic}
\end{algorithm}

\section{Configuration}
\label{app:config}

Table~\ref{tab:hparams} lists the configuration of the final method. Model weights are
never updated; the values govern the memory-editing loop and the per-task execution
budget. Constants are pre-registered: a change defines a new condition, not a tuning.

\begin{table}[h]
\caption{\textbf{Configuration of EMMA.}}
\label{tab:hparams}
\begin{center}
\scriptsize
\renewcommand{\arraystretch}{1.1}
\resizebox{\linewidth}{!}{%
\begin{tabular}{ll}
\toprule
\textbf{Setting} & \textbf{Value} \\
\midrule
Training passes & 1, fixed task order \\
Training gate & KramaBench: Opus~4.5 judge; AppWorld: unit tests; BBEH: exact match \\
Revisions per task $K$ / log inspections per revision & 5 ($\le 6$ executions) / 3 \\
Refusal feedback rounds before forced \act{no\_change} & 3 \\
End-of-task controller-memory entries & $\le 2$ \\
Pool caps: agents (incl.\ seed) / skills; chars per prompt / skill & 10 / 40; 6{,}000 / 3{,}000 \\
Prompt-edit line retention / minimum specialization & 0.6 / 150 chars \\
Task-text leak span & 15 words \\
Level rotation & no level in $> 2$ consecutive failed cycles \\
Controller-memory loading budget & 15 entries / 12{,}000 chars \\
Rounds per agent per step $R$ (round cap = overflow) & 30 \\
Plan steps $K_\Pi$ (final answer step always executed) / re-plan attempts & 15 / 3 \\
Executor timeout / child memory limit / captured output & 600\,s / 24\,GiB / 2\,MB per stream \\
Controller / sub-agent reasoning & high (adaptive thinking on Claude~5) / low \\
Transport retry ladder (KramaBench; BBEH, AppWorld) & $12 \times 10$\,s; $60 \times 10$\,s \\
\bottomrule
\end{tabular}}
\end{center}
\end{table}

\textbf{Action vocabulary.} Table~\ref{tab:actions} lists the reflector's actions.
The guards of Section~\ref{sec:method-guards} are checked in the following order:
scope--action compatibility; change of strategy (escalation after repeated overflow,
then level rotation); no leakage (answer in artifacts or directives, verbatim task
text); prompt preservation (participant-only edit, line retention, length); bounded
memory growth (skill bounds, pool cap, duplicate, locator and label-only agents,
dataset or run paths in an artifact).

\begin{table}[h]
\caption{\textbf{Reflector action vocabulary.}}
\label{tab:actions}
\begin{center}
\scriptsize
\renewcommand{\arraystretch}{1.1}
\begin{tabular}{llll}
\toprule
\textbf{Action} & \textbf{Level} & \textbf{Durable} & \textbf{Payload} \\
\midrule
\act{no\_change} & none & no & the failure is not addressable by the program \\
\act{revise\_plan} & plan directive & this task only & a directive the next attempt's planner must follow \\
\act{memory\_note} & controller memory & yes & topic, one to three sentences, generality label \\
\act{refine\_prompt} & agent prompt & yes & the complete edited prompt of a participating agent \\
\act{revise\_skill} / \act{create\_skill} & skill pool & yes & skill text and its ``when to use'' description \\
\act{create\_agent} & agent pool & yes & name, description, specialization (inherits the generalist) \\
\bottomrule
\end{tabular}
\end{center}
\end{table}

\section{Benchmarks and Splits}
\label{app:details}

Every split is disjoint: no training task appears in any evaluation set, and every
system is trained and evaluated on the same split. Split membership is generated once
from pinned sources and loaded through a fingerprint-verifying loader that refuses to
run on a mismatch. Table~\ref{tab:splits} summarizes the splits.

\begin{table}[h]
\caption{\textbf{Benchmark splits.}}
\label{tab:splits}
\begin{center}
\scriptsize
\begin{tabular}{lrrrl}
\toprule
\textbf{Benchmark} & \textbf{Total} & \textbf{Train} & \textbf{Eval} & \textbf{Split construction} \\
\midrule
KramaBench & 104 & 28 & 76  & deterministic, stratified by domain and difficulty \\
AppWorld   & --- & 90 & 168 (normal) / 417 (challenge) & upstream splits, unmodified \\
BBEH-mini  & 460 & 92 & 368 & seed 42, 4 train / 16 eval per category \\
\bottomrule
\end{tabular}
\end{center}
\end{table}

\textbf{KramaBench.} The benchmark supplies 104 tasks across six domains and defines
no split. Within each domain, tasks are separated into the \texttt{easy} and
\texttt{hard} groups the benchmark labels, sorted by identifier, and every fifth task
is assigned to training. The rule is index-based, requires no random seed, and
stratifies over both domain and difficulty; the per-domain counts are given in
Table~\ref{tab:kb-split}.

\begin{table}[h]
\caption{\textbf{KramaBench per-domain split.}}
\label{tab:kb-split}
\begin{center}
\scriptsize
\begin{tabular}{lrrrrrrr}
\toprule
 & \textbf{archeo.} & \textbf{astro.} & \textbf{biomed.} & \textbf{environ.} & \textbf{legal} & \textbf{wildfire} & \textbf{total} \\
\midrule
Train & 4 & 4 & 3 & 5  & 7  & 5  & 28 \\
Eval  & 8 & 8 & 6 & 15 & 23 & 16 & 76 \\
\bottomrule
\end{tabular}
\end{center}
\end{table}

\textbf{AppWorld.} We use the benchmark's own \texttt{test\_normal} and
\texttt{test\_challenge} splits without modification and train on its 90-task
training split. TGC and SGC are computed by the benchmark's state-based unit tests.

\textbf{BBEH-mini.} 23 categories $\times$ 20 examples $= 460$, split per category into
4 training and 16 evaluation examples under seed 42, so all categories appear on both
sides. Because every category contributes exactly 16 evaluation examples, macro and
micro accuracy coincide on this split.

\section{Baseline Methods}
\label{app:baselines}

All four baselines in the main comparison learn persistent artifacts from training
feedback and freeze them for evaluation. DACode, smolagents DeepResearch and Reflexion
appear only in the ablation of Section~\ref{sec:exp-ablation}.

\textbf{ACE}~\citep{ace} grows a playbook of identified bullets for a fixed ReAct agent
with code execution. A generator solves the task with the playbook in context, a
reflector tags bullets as helpful or harmful, and a curator applies incremental edits;
the playbook with the best validation accuracy is frozen. Feedback is answer
correctness on KramaBench and, on AppWorld, the no-ground-truth mode in which the
reflector sees only unit-test pass/fail and the trace.

\textbf{GEPA}~\citep{gepa} evolves the prompt texts of a compound system. Candidates are
evaluated on training mini-batches; a reflection model reads the traces and rewrites
one component; a Pareto front over validation tasks (drawn from the training split)
selects survivors. Feedback is the benchmark metric plus the execution traces.

\textbf{EvoMAS}~\citep{evomas} evolves complete multi-agent configurations (topology,
prompts, models, tools) by meta-model selection, mutation and crossover for three
steps, keeping successful configurations in an experience memory. In the main
comparison the reward is the dataset evaluator (oracle mode) minus a token-and-latency
cost term, and the best configuration and memory are frozen. The separate EvoMAS arm
of Table~\ref{tab:ablation} adapts on test tasks through self-reflection without
ground truth.

\textbf{ADAS}~\citep{adas} has a meta-agent write complete agent programs in code,
evaluates each on the training split, and keeps an archive that conditions the next
proposal, with a budget matched to GEPA. The best archived program is frozen; where no
discovered program beat the hand-written seed (DeepSeek~v3.2 and Kimi-K2.5 on
AppWorld), the seed is reported.

\subsection{Update Constraints in Learned Baselines}
\label{app:guard-compare}

Table~\ref{tab:guard-compare} compares how each method constrains a single update, as
described in the respective papers~\citep{ace,gepa,evomas,adas}. ``Not described''
means that the paper specifies no such mechanism. The two multi-agent search methods
leave both the locus and the magnitude of a change to the backbone: ADAS checks a
proposal only for novelty (two self-reflection steps) and executability (up to five
debugging rounds), and EvoMAS restricts a mutation to one component type (prompts,
models, tools or topology) but not how much of that component changes. Neither types
the update by a diagnosed failure, bounds the resulting system, or checks for
leakage. ACE's incremental delta bullets with embedding de-duplication are the
closest analogue of an edit guard, but act on a single context level.

\begin{table}[h]
\caption{\textbf{How a single update is constrained.}}
\label{tab:guard-compare}
\begin{center}
\scriptsize
\renewcommand{\arraystretch}{1.15}
\resizebox{\linewidth}{!}{%
\begin{tabular}{llllll}
\toprule
\textbf{Method} & \textbf{Unit of update} & \textbf{Where to edit} & \textbf{Bound on edit} & \textbf{Leakage check} & \textbf{Acceptance} \\
\midrule
ACE & delta bullets in one playbook & single level & incremental; de-duplication & not described & best-validation playbook \\
GEPA & one component prompt & round-robin scheduler & none (full rewrite) & not described & minibatch improvement \\
EvoMAS & whole configuration & meta-model & one component type per mutation & not described & evaluator reward $-$ cost \\
ADAS & whole agent program (code) & meta-agent & none (novelty, debugging) & not described & archived after evaluation \\
\midrule
EMMA & one entry at one level & diagnosed scope, as in Table~\ref{tab:scopes} & prompt preservation, bounded growth, change of strategy & answer and task text & re-run on a fresh environment \\
\bottomrule
\end{tabular}}
\end{center}
\end{table}

\section{Reflector Prompt}
\label{app:reflector}

The prompt below drives the reflection step. Run-time inputs appear in angle brackets,
pre-registered constants are filled in, and the two conditional sections (overflow
evidence; escalation) are shown where they appear when present.

\begin{lstlisting}[style=artifact]
A training task FAILED its correctness gate. Diagnose the failure from the execution trajectory, decide which LEVEL of the program is responsible, and make exactly ONE improvement before the task is re-run on a fresh environment.

## The program you are improving
The program is `P + router + {skills}`, plus the ability to add agents. Each level has a home for lessons; a lesson written at the wrong level either never fires again or poisons unrelated tasks.
- ROUTER (controller memory): how the controller decomposes a task and dispatches subtasks to agents and skills. Home for lessons about decomposition, sequencing, which agent or skill to use, and verification steps in the plan. Action: memory_note.
- AGENT PROGRAM P (an agent's system prompt): that agent's general working method -- how it reads a subtask, plans, executes, checks and reports -- independent of any topic. Home for behaviors that would recur on ANY subtask the agent handles. Action: refine_prompt (an EDIT of the existing prompt: keep what works, add or amend a few lines).
- SKILLS (procedural modules the router assigns to an agent per step): a self-contained method for a recognizable CLASS of sub-problems, written so it transfers to unseen instances of that class -- an algorithm, a rule system, a checklist of pitfalls, a format convention. Actions: revise_skill (a listed skill covers this class but is wrong or incomplete), create_skill (no skill covers it).
- AGENTS (roles): a new agent only when a kind of work needs its own working method and context that does not fit any existing agent's P and cannot be expressed as a skill. Action: create_agent (rare; the pool is capped at 10).
Non-durable outcomes: revise_plan -- a directive that only the next attempt of THIS task follows (a one-off slip, or a different decomposition to try); no_change -- the failure is not addressable by the program (e.g. the reference or gate looks wrong).

## Procedure -- follow in order and fill every field of the output
1. LOCALIZE. From the plan, the action trace and the evaluator digest, identify the step and agent where the outcome went wrong, or conclude that the plan itself was wrong. Call inspect_step only when the code and digest are not diagnostic by themselves.
2. CLASSIFY the scope -- exactly one:
   - execution_slip: the agent knew the right method and slipped once (misread, off-by-one, typo); would not recur systematically. -> revise_plan or no_change
   - agent_behavior: a general working habit failed (did not verify, did not read the whole input, answered before computing, ignored the requested format, stopped early); recurs across topics. -> refine_prompt
   - procedure: the method for this class of sub-problem was wrong or missing (a domain algorithm, a rule system, a convention). -> revise_skill if a listed skill covers the class, else create_skill
   - missing_capability: a kind of work nobody in the pool can do. -> create_skill (preferred) or create_agent (only if it needs its own working method and context)
   - routing: the agents did their subtasks correctly but the decomposition or dispatch was wrong (wrong agent, missing step, no verification step, a relevant skill not assigned). -> memory_note or revise_plan
3. GENERALIZE. Phrase the change so it applies to every future task of the same KIND. It must not mention this task's entities, values, names, options or answer, and must not quote the task text. Prefer principles and procedures over facts about this instance.
4. MINIMIZE. The smallest change, at the lowest level that actually owns the lesson. One action.

## Hard rules (violations are refused and you must redo the action)
- Neither artifacts nor plan directives may contain this task's answer; artifacts never contain verbatim spans of the task text. A directive that tells the next attempt what the answer is teaches nothing and is refused.
- refine_prompt is an edit of an agent that PARTICIPATED in this attempt (their complete prompts are shown below): at least 60% of the existing prompt's lines must survive; add or amend a few lines; the prompt stays under 6000 characters.
- Level rotation: no level may be used in more than 2 consecutive cycles of one chain. If the chain history shows the last 2 cycles at the same level and the task still fails, that level is refused now -- move to another level, or use revise_plan / no_change.
- Skills stay under 3000 characters, are procedural and generic to their class, and come with a one-line index description saying WHEN to use the skill.
- create_agent is refused when the pool holds 10 agents or the name exists; prefer a skill.
- The declared scope and the action must match the table above.

## Inheritance and targeting (v1.4.6)
- Every created agent RUNS with the generalist's CURRENT prompt followed by its own specialization. The prompt shown below for a created agent is ONLY its specialization; the generalist's prompt is shown read-only when the generalist did not participate.
- A lesson about a general working habit (verifying, reading the whole input, computing before answering, the requested output format, persisting when a first search finds nothing, which source to trust) belongs in the GENERALIST's prompt -- refine_prompt on `generalist` -- because inheritance carries it to every specialist. This is allowed even when the generalist did not participate, provided the failing agent inherits from it.
- A lesson specific to a specialist's role belongs in THAT specialist's specialization -- refine_prompt on the agent that ran the failing step. Never write a general rule into a specialization; never write a role-specific procedure into the generalist.
- create_agent under inheritance: details.system_prompt is ONLY the specialization (the generalist's method is inherited automatically).

## Task
<task text>

## Evaluator digest (why the gate failed this attempt)
<EVALUATOR DIGEST>

## Plan (this attempt)
<PLAN OF THIS ATTEMPT>

## Action trace (per step, per round: the code issued -- environment output deliberately omitted)
<ACTION TRACE>

## Chain history (this task's prior revision cycles)
<PRIOR REVISION CYCLES OF THIS TASK>

## Agent pool (<n>/10)
<AGENT POOL SUMMARY>

## Current prompts of the agents that participated (complete text -- edit, do not rewrite)
<FULL PROMPTS OF PARTICIPATING AGENTS>

## Skills index (<k>/40; the router assigns these per step)
<SKILL INDEX>

## Skills injected in this attempt (complete text)
<FULL TEXT OF INJECTED SKILLS>

## OVERFLOW evidence (v1.4.8) -- steps that exhausted an agent in this attempt
<one line per overflowed step: step * agent * kind * rounds used * decomposition outcome; subtask; partial trace>
With this evidence, create_agent is admissible under scope routing as well as missing_capability: a kind of work that repeatedly exhausts one agent's rounds or context needs its own agent with a narrower context, and when the same step overflowed in two consecutive attempts create_agent is the EXPECTED structural response. A new agent must be a genuinely narrower capability: not a duplicate of an existing agent, never a "find/locate the files" role (the inputs are given -- no agent searches for them), and its prompt must name no paths, datasets or runs.

## ESCALATION -- structural action REQUIRED
Step(s) <step numbers> overflowed in two consecutive attempts, even after mid-task overflow decomposition. A prompt or skill edit is not acceptable here: you MUST choose a structural action -- create_agent (scope missing_capability) or revise_plan (scope routing / execution_slip) that changes how the work is split.

## Drill-down (optional)
You may call inspect_step(step_index, round_start, round_end) to see the full execution log (code + stdout + stderr) for specific rounds. Budget: 3 inspections this cycle.

## Output -- JSON only
~~~json
{
  "diagnosis": {
    "failing_step": <step number or null>,
    "failing_agent": "<agent name or null>",
    "what_went_wrong": "one or two sentences",
    "scope": "execution_slip | agent_behavior | procedure | missing_capability | routing",
    "recurrence": "why this would or would not recur on other tasks of this kind"
  },
  "generality_check": "confirm in one sentence that the change names no task-specific entities, values or answers",
  "action": "<action>",
  "details": {...}
}
~~~
Actions and their details:
- no_change: {}
- revise_plan: {"plan_directive": "a concrete directive the next attempt's planner must follow (this task only)"}
- memory_note: {"topic": "short topic", "content": "1-3 sentences: the routing/decomposition rule", "generality": "general | domain_specific"}
- refine_prompt: {"agent": "existing agent name", "new_prompt": "the COMPLETE edited prompt text"}
- revise_skill: {"skill_name": "existing-skill-name", "new_content": "the complete revised skill (markdown)", "description": "one line: when to use it"}
- create_skill: {"skill_name": "kebab-case-name", "content": "the skill (markdown)", "description": "one line: when to use it", "agent": "existing agent that should use it"}
- create_agent: {"name": "snake_case_name", "description": "one line: what kind of work it does", "system_prompt": "its general working method"}
\end{lstlisting}

\section{What Training Produces: Learned Artifacts}
\label{app:artifacts}

In the Sonnet~5 BBEH run, one failure chain populated all three levels
. A failed training zebra puzzle was classified as
\scope{procedure} and produced the skill \texttt{zebra-csp-encode-and-verify}; the
re-run split the puzzle across many steps and overflowed, and with this evidence the
reflector created \texttt{csp\_solver}, a constraint-satisfaction specialist inheriting
the generalist. The task passed on attempt 5, and the end-of-task update wrote a
routing rule: \emph{solve large zebra puzzles in a single \texttt{csp\_solver} step}.
The next three zebra training tasks passed on the first attempt. On a held-out puzzle
with 52 clues, the controller loaded 8 of 101 entries, dispatched one
\texttt{csp\_solver} step with the skill, and obtained a solution verified unique by a
blocking re-solve in 9 rounds. None of the three entries names an entity or answer of
the training puzzles.

The artifacts below are the frozen final memory of this run and the corresponding
evaluation log.

\textbf{Controller-memory entry} (type \texttt{workflow\_pattern}, generality
\texttt{domain\_specific}, domain \texttt{bbeh\_zebra\_puzzles}). Two later entries
restate the rule, noting that it ``succeeded on first attempt with no revision
needed'' and ``scales fine even with 15+ categories per position''.
\begin{lstlisting}[style=artifact]
For large zebra/logic-grid puzzles (100+ clues), route to a single csp_solver step that transcribes, encodes, and solves all clues in one continuous pass rather than splitting into transcription/collision-check/build/diagnose/resolve sub-steps -- checkpoint files frequently fail to persist across steps, forcing costly re-derivation and wasted rounds on unsat-core chasing before eventually re-solving from scratch anyway.
\end{lstlisting}

\textbf{Agent specialization} \texttt{csp\_solver} (inherits \texttt{generalist}; 31
recorded invocations), appended to the generalist's current prompt at dispatch. Its
description reads: ``Handles large constraint-satisfaction problems (e.g.\ many-clue
logic-grid/zebra puzzles): incremental model construction, solving, and
unsat-core-based debugging, split into small checkpointed batches so no single call
needs to do transcription+encoding+solving+diagnosis all at once.''
\begin{lstlisting}[style=artifact]
You specialize in building and solving LARGE constraint-satisfaction problems (CSP), such as logic-grid/zebra puzzles with many categories and many numbered clues, using OR-Tools CP-SAT or equivalent.

Working method:

1. Never attempt to transcribe the source, encode every constraint, solve, AND debug infeasibility inside a single call. Split the work yourself into checkpointed phases and persist state to files between phases: (a) transcribe categories/values and clue text once, save verbatim; (b) build one small set of well-tested helper functions per relational pattern (identity, negation, adjacency, ordered-between, at-end, fixed-position) and unit-test each on a tiny toy example before using it; (c) encode constraints in fixed-size batches (e.g. 25-40 clues at a time), saving the partial model/constraint list to a file after each batch so later batches or later steps can resume without re-deriving; (d) only after all batches are encoded, solve once.

2. Track your round budget explicitly. If you are more than half your round allowance into a phase without finishing it, stop expanding scope, write out your current partial artifact (encoded constraints so far, or the current unsat core) to a file, and report exactly what is done and what remains, so a follow-up call can continue from that checkpoint instead of restarting.

3. If the full model reports INFEASIBLE, do not iteratively hand-drop clues or hand-derive chains of inequalities as a substitute for solving. Instead: extract the unsat core programmatically (e.g. CP-SAT SufficientAssumptionsForInfeasibility) exactly once, then for every clue in that core re-derive its encoding independently from the literal clue text, diff it against the original, fix any mismatch, and re-solve the FULL model. Do at most one such re-audit cycle per call; if still infeasible after that, checkpoint the core and mismatches found and stop rather than manually enumerating chains of hand arithmetic.

4. Before reporting a solution as final, re-solve with the found assignment excluded to confirm uniqueness. Report clearly which phase you completed, what artifact you saved, and what the next call should resume with.
\end{lstlisting}

\textbf{Procedural skill} \texttt{zebra-csp-encode-and-verify}, indexed as ``Use when a
subtask requires encoding a multi-category zebra/logic-grid puzzle (many numbered
clues, many categories, positions in a row/grid) into a CSP and verifying it''.
\begin{lstlisting}[style=artifact]
## Zebra/logic-grid CSP encoding and verification (CP-SAT or equivalent)

1. UNIQUE NAMES PER CATEGORY: give every value a name unique across ALL categories (e.g. prefix with category), never reuse a bare word like 'biking' in two different categories (hobby vs activity). Category-value collisions silently corrupt constraints and are a common source of false INFEASIBLE.

2. PIN DOWN PRIMITIVES ONCE, THEN REUSE: before encoding any clue, write exactly one helper function per relational pattern actually used by the clues (identity/'is the', negation/'is not', 'immediately left/right of' as pos(A)+1==pos(B), 'next to'/'adjacent' as |pos(A)-pos(B)|==1, 'X is somewhere in between A and B in that order' as pos(A)<pos(X)<pos(B), 'at one of the ends', fixed absolute position references like 'the person at position K'). Test each helper on a tiny 2-3 variable toy example that has a known expected SAT/UNSAT outcome before applying it to all clues.

3. ENCODE ALL CLUES IN ONE PASS, VERBATIM: extract literal clue text once, then encode every clue directly against its exact wording using only the verified helpers from step 2. Do not re-interpret or 'improve' phrasing case by case.

4. TREAT A FIRST INFEASIBLE AS A BUG SIGNAL, NOT A PUZZLE FACT: these puzzles are constructed to be satisfiable. If the full model reports INFEASIBLE, do NOT start dropping/relaxing individual clues as if some clue is 'wrong' -- that treats an assumed defect in the puzzle as ground truth. Instead: (a) recompute the unsat core, (b) for every clue in the core, re-derive its encoding from the literal text completely independently (fresh helper call, fresh mental parse) and diff against the original encoding, (c) specifically re-check direction (which side is 'immediately left' vs 'right'), identity vs adjacency conflation, and any off-by-one in absolute position indices. Fix any mismatch found and re-solve the FULL model (not a dropped-clue variant). Only conclude a clue is genuinely unsatisfiable with the rest after this re-audit still reproduces the same minimal core with all encodings independently re-verified as literal-correct.

5. VERIFY UNIQUENESS ONLY AFTER SAT: once SAT, re-solve with the found full assignment excluded (add a blocking clause) to confirm UNSAT (unique solution) before reporting any derived attribute as the final answer. If a second solution exists, the model is underconstrained -- check for an unencoded or misencoded clue.

6. BUDGET DISCIPLINE FOR LARGE PUZZLES: encoding 100+ clues and solving is one continuous task; do not spend rounds re-deriving the same helper twice. If you are more than half your round budget in without a completed, verified SAT solve, stop broad exploration (e.g. stop trying clue-drop variants) and instead do the targeted re-audit in step 4 on the actual unsat core -- this converges faster than guessing which clue to relax.
\end{lstlisting}

\textbf{Evaluation trace} on held-out task \texttt{bbeh\_zebra\_puzzles-00}.
\begin{lstlisting}[style=artifact]
[v143 Memory] controller selected 8/101 entries

[Controller] Decomposing task: bbeh_zebra_puzzles-00

[Controller] Plan: 2 steps

[Controller] Step 1: Solve this 8-position zebra/logic-grid puzzle. There are 8 people in positions 1... (agent: csp_solver, skills: ['zebra-csp-encode-and-verify'])

[Agent csp_solver] Completed in 9 rounds

[Controller] Terminating: The csp_solver already produced a fully verified, unique solution confirming the person who likes beavers is at position 6. All 52 clues were checked and uniqueness was verified by blocking and ...

[Controller] Task completed in 82.3s

  [bbeh_zebra_puzzles-00] correct=1
\end{lstlisting}

Table~\ref{tab:three-levels} summarizes how the three levels cooperate on this task.

\begin{table}[h]
\caption{\textbf{Three memory levels on a held-out zebra puzzle.}}
\label{tab:three-levels}
\begin{center}
\scriptsize
\renewcommand{\arraystretch}{1.1}
\begin{tabular}{p{1.9cm}p{3.1cm}p{2.5cm}p{3.9cm}}
\toprule
\textbf{Level} & \textbf{Entry} & \textbf{Responsibility} & \textbf{Observed evidence} \\
\midrule
Controller memory & single \texttt{csp\_solver} step, one continuous pass & decomposition and dispatch & the plan uses one solving step and repeats the routing rule \\
Agent configuration & \texttt{csp\_solver} specialization & model construction, debugging, verification & the specialist completes in 9 rounds and reports a verified solution \\
Procedural skill & \texttt{zebra-csp-encode-}\allowbreak\texttt{and-verify} & clue encoding and correctness checks & the skill is assigned; the trace reports clue checks and a blocking re-solve for uniqueness \\
\bottomrule
\end{tabular}
\end{center}
\end{table}

\end{document}